\documentclass{conference}
\usepackage{graphicx}
\usepackage{amssymb}

\begin{document}
\begin{frontmatter}
\title{THE VOLUME CAPTURE IN STRUCTURES WITH VARIABLE CURVATURE}

\author{Gennady V. Kovalev}
\author{}
\address{School of Mathematics \\University of Minnesota, Minneapolis, MN 55455,USA}

\begin{abstract}
The volume capture in classical relativistic mechanics is considered as a scattering process for the high energy charged particles  in a field with no central or mirror symmetry.  The parameters of volume capture for potentials with smooth variable curvature are received and analyzed. 
\end{abstract}

\end{frontmatter}

\section{\label{sec:01} Introduction}
It is known that the symmetry of straight or bent crystal with a constant curvature does not allow the volume capture of particles in the channeling motion in Hamiltonian mechanics. In quantum mechanics, however, there is a possibility of the volume capture due to an exponentially small tunneling through the potential barrier, but we will not consider this effect here. We will also not touch the incoherent scattering which is the major cause of the volume capture ~\cite{tar98,bck} in straight and circular bent crystals with constant curvature as well as the normal dechanneling. Our aim is to consider the volume capture, as a pure classical effect, which becomes possible due to curvature variation. In some works ~\cite{tar98,bck} it is called the 'gradient capture', and it was pointed out first in the computer simulation ~\cite{mannami_1988}. Here we describe some general features and relations of volume capture as a classical effect of relativistic electrodynamics starting from classical potential scattering.
 
\section{\label{sec:02} Classical Scattering and Channeling in Uniform Field with Curved Boundary}
\begin{figure}
	\centering
		\includegraphics{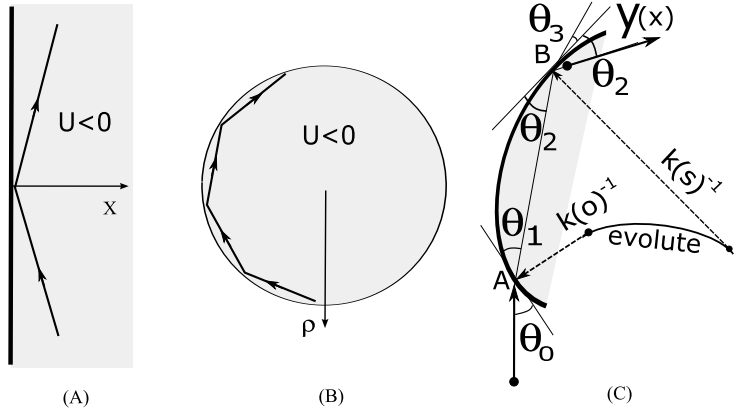}
	\caption{(A)Reflection from a wall (B)Channeling inside circular cylinder (C)Volume capture  }
	\label{fig:CurveSurface}
\end{figure}

As a physical phenomenon, the channeling can be considered as a finite motion in one or two directions (oscillations or more complicate finite motion) and free motion in third, generally speaking, curvilinear direction \cite{kov06_1}. Such motion can be imposed by a force field in a neighborhood of smooth curve or plane in 3D space. The curvature of the curve or plane surface can also play the role of a force field. If the electric force in a neighborhood of a plane with zero curvature is applied in one direction, it is not enough to localize a particle near the surface. From this point of view, the simple reflection of particles from a flat potential ($U(x)<0$ if $x>0$ and $U(x)=0$ if $x<0$, Fig.~\ref{fig:CurveSurface} (A)) is not a channeling, because there is no finite motion in direction perpendicular to the plane. But if the potential is confined in a circular cylinder, ($U(\rho)<0$ if $\rho<R$ and $U(\rho)=0$ if $\rho>R$) or sphere, the classical consecutive reflection looks like oscillation in a neighborhood near the potential wall and can be called the 'surface channeling' (see Fig.~\ref{fig:CurveSurface} (B)).     

However, if the initial coordinates of the particle is located outside the cylindrical well,  any trajectory is a symmetrically refracted on entering and leaving the potential well (see e.g. \cite{ll1}, $\S$14,18). Thus the scattering trajectories of a particle by a circular cylindrical or spherical potential does not include the channeling trajectories. This is not the case, if the particles are scattered by a potential with variable curvature, as it will be shown below.

We shall first consider the deflection of a relativistic  
particle of mass $m$ and energy $E$ crossing the boundary of a uniform field $U_1$ which occupies the right side of a curve $y=Y(x)$, Fig.~\ref{fig:CurveSurface} (C), with the curvature $k(s)$, where $s$ is the natural parameter. The path of a particle is refracted at the angles $\Theta_0$  and $\Theta_2$  to the tangent lines, respectively at point A and B. The tangent continuity of a particle's momentum gives the relativistic relations between pair of angles  $\Theta_0$, $\Theta_1$  and   $\Theta_2$, $\Theta_3$  :

\begin{eqnarray}	
(\phi+1)sin^2(\Theta_1) = \phi+sin^2(\Theta_0),\;\label{Refraction1}\\
(\phi+1)sin^2(\Theta_2) = \phi+sin^2(\Theta_3),
\label{Refraction2}
\end{eqnarray}

where the 
\begin{eqnarray}	
\phi=\frac{2E|U_1|}{E^2-m^2c^4}
\label{LindhardAngle}
\end{eqnarray}

is the square of Lindhard's angle. In deriving (\ref{Refraction1}),(\ref{Refraction2}), we made one assumption, $|U_1|<<m \leq E$, which is true for relativistic as well as nonrelativistic particles. 
When the curvature $k$ of boundary is constant (circular symmetry), the equality $\Theta_1=\Theta_2$  holds, and the equations (\ref{Refraction1}),(\ref{Refraction2}) give $\Theta_0=\Theta_3$ . 
Here is no volume capture. Another situation happens when the curvature   at the second point of crossing B is less than curvature   at the point A. In this case $\Theta_2$  becomes smaller than $\Theta_1$, but $\Theta_3$  is always smaller than $\Theta_2$ if the potential $U_1$ is negative inside the boundary. When the curvature at B continuously decreases, the condition $\Theta_3=0$ takes place and at this point the second eq. (\ref{Refraction2}) gives: 
\begin{eqnarray}	
\Theta_{2cr} = arcsin(\sqrt{\frac{\phi}{(\phi+1)}}).
\label{crAngle2}
\end{eqnarray}
The particle undergoes total internal refraction at point B, if $\Theta_2 <\Theta_{2cr}$. It will be captured in channeling mode if the curvature remains the same or decreases further along the boundary of potential. 
The integral curvature between points A and B defines the relation between the angles $\Theta_1$ and $\Theta_2$:

\begin{eqnarray}	
\Theta_2+\Theta_1 = \int_{s_A}^{s_B} k(s) ds.
\label{AngleRelationGeneral}
\end{eqnarray} 
In this form it generalizes the way of calculating the angle between two tangent lines  to the two separate points on the smooth curve. The the radius vectors of the centers of curvatures are tangent to the evolute of the smooth curve (see Fig.\ref{fig:CurveSurface}). The evolute curve can not be smooth, in general.  

One of the most important applications of the formulae derived above is to reverse the scattering problem and put the following question. What would be the incident angle $\Theta_0 $, if the internal glancing angle $\Theta_2 \leq \Theta_{2cr} $? 
From (\ref{Refraction1}), we can receive

\begin{eqnarray}	
sin(\Theta_{0}) =\pm \sqrt{(\phi+1)sin^2(\Theta_{1})-\phi}.
\label{Angle0}
\end{eqnarray}

Using the critical angle $\Theta_{2cr}$ (\ref{crAngle2}) and the relation (\ref{AngleRelationGeneral}), we obtain an expression for critical angle $\Theta_{1cr}$:

\begin{eqnarray}	
\Theta_{1cr} =\int_{s_A}^{s_B} k(s) ds -\Theta_{2cr}.
\label{crAngle1}
\end{eqnarray}
Now, the critical angle  $\Theta_{0cr}$ becomes
\begin{eqnarray}	
sin(\Theta_{0cr}) =\pm \sqrt{(\phi+1)sin^2(\int_{s_A}^{s_B} k(s) ds- arcsin(\sqrt{\frac{\phi}{(\phi+1)}}))-\phi}.
\label{crAngle0}
\end{eqnarray}

\begin{figure}
	\centering
		\includegraphics{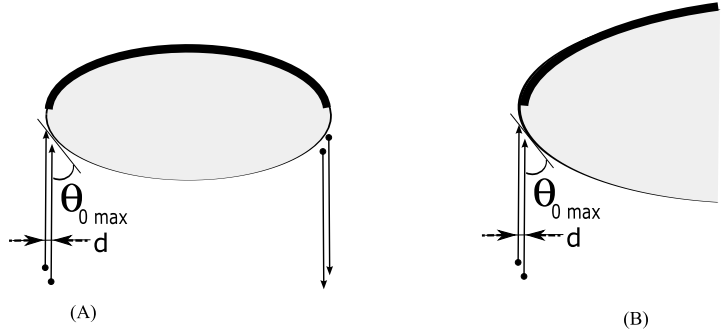}
	\caption{The volume capture in (A)elliptic and in (B)parabolic cylindrical wells }
	\label{fig:Ellipse}
\end{figure}

 For an infinite motion, such as that considered here, if we increase  the impact parameter $d$  relatively the edge of the potential (see Fig.\ref{fig:Ellipse}), the angle $\Theta_{0}$ will vary from $0$ to $\Theta_{0cr}$. Inside this angle the relativistic particles will be captured in channeling mode by 'volume' effect.  Expressing the impact parameter  in terms of curvature  gives 

\begin{eqnarray}	
d_{max}= k(0)^{-1}-k(s_A)^{-1}cos(\Theta_{0cr}),
\label{ImpactParamG}
\end{eqnarray}
here $k(0)$ is the curvature at the natural parameter where $\Theta_{0cr}=0$. 
This impact parameter can be used for the calculations of the channeling cross-section.
For the small angles $\Theta_{0cr}$ :
\begin{eqnarray}	
d_{max}= k(0)^{-1}-k(s_A)^{-1}+\frac {(\Theta_{0cr})^2}{2k(s_A)}.
\label{ImpactParamAproc1}
\end{eqnarray}
If the curvature's variation is small $k(x_A)^{-1}\cong k(0)^{-1}- \frac{k(0)^{'}}{k(0)}s_A$, then
\begin{eqnarray}	
d_{max}= \frac{k(0)^{'}}{k(0)}s_A+\frac {(\Theta_{0cr})^2}{2k(0)}.
\label{ImpactParamAproc2}
\end{eqnarray}

This situation, strictly speaking, is different from classical scattering on symmetrical potentials considered in many textbooks (see e.g. \cite{ll1}, $\S$14,18), because the particles can move along the boundary and be reflected back (see Fig.\ref{fig:Ellipse}). 

If there is no curvature variation, then $k(0)^{'}=0, \Theta_{0cr}=0$ (potential with constant curvature) and the parameter of cross section for volume capture $d_{max}=0$.

Since the all angles $\Theta_{0}$,$\Theta_{1}$,$\Theta_{2}$ are very small, the eq.(\ref{crAngle0}) can be written in the following form

\begin{eqnarray}	
\Theta_{0cr} =\sqrt{(\phi+1)(\int_{s_A}^{s_B} k(s)ds - \sqrt{\frac{\phi}{(\phi+1)}})^2-\phi}.
\label{crAngleM}
\end{eqnarray}
\section{\label{sec:03} Efficiency of Capture}
 
Now we consider the volume capture of the particles by potential boundary with parabolic shape $y=ax^2$. The curvature of parabola is

\begin{eqnarray}	
k(x) =\frac{2a}{(1+4a^2x^2)^{3/2}}.
\label{ParebolaCurvature}
\end{eqnarray} 

To use the curvature in calculation of critical angles (\ref{crAngle0}), we need the natural parameter $s$, which is a distance along the parabolic curve
 
\begin{eqnarray}	
s(x) =\int_0^x \sqrt{1+(\frac{dy(t)}{dt})^2}dt,
\label{NaturalParameter}
\end{eqnarray} 
or in differential form
\begin{eqnarray}	
\frac{ds}{dx} =\sqrt{1+(\frac{dy(x)}{dx})^2}.
\label{difNaturalParameter}
\end{eqnarray}

Using the differential form (\ref{difNaturalParameter}), the integral in angle relation (\ref{AngleRelationGeneral}) can be presented by

\begin{eqnarray}	
\int_{s_A}^{s_B} k(s) ds =\int_{x_A}^{x_B} k(x) \frac{ds}{dx} dx,
\label{AngleRelationParametric}
\end{eqnarray} 
and taken between x-coordinates of points A and B on the boundary. The result of its calculation is

\begin{eqnarray}	
\Theta_2+\Theta_1 = arctan(2ax_B)-arctan(2ax_A).
\label{AngleRelationParabola}
\end{eqnarray} 

In this formula we don't know the mutual location of points $x_A$ and $x_B$. By assuming that the particle in uniform field moves in a straight line between the points A and B, the angles $\Theta_1$, $\Theta_2$ can be calculated separately:

\begin{eqnarray}	
\Theta_1 = arctan(ax_B+ax_A)-arctan(2ax_A),\label{AnglesSeparate1}\\
\Theta_2 = -arctan(ax_B+ax_A)+arctan(2ax_B).
\label{AnglesSeparate2}
\end{eqnarray}

\begin{figure}
	\centering
		\includegraphics{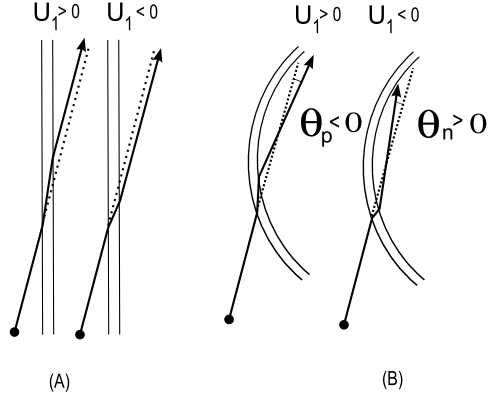}
	\caption{The trajectories of nonchanneled particles are shifted in straight crystals (A). In bent crystals they are rotated toward the direction of bending for positive particles and in the opposite direction for negative particles. If the counterclockwise direction is considered as a positive then $\Theta_p <\Theta_n$.  }
	\label{fig:ShiftRotation}
\end{figure}

 The angle $\Theta_1$, it should be noted, always greater than $\Theta_2$ and  monotonically increases with $ x_B$, as $x_A,$ is constant. The angle $\Theta_2$ decreases as $ x_B \rightarrow \inf$ and $x_A$ stays constant.
Since (\ref{AnglesSeparate1}),(\ref{AnglesSeparate2}) shows that the difference
  
\begin{eqnarray}	
\Theta_1 - \Theta_2= 2 arctan(ax_B+ax_A)-arctan(2ax_A)-arctan(2ax_B),
\label{AnglesDifference}
\end{eqnarray} 
 
can never change sign if $x_B > x_A$.  Because $\Theta_2$ decreases sharply, it can reach the Lindhard's angle (\ref{crAngle2}) and  conditions of volume capture (\ref{crAngle0}) can be satisfied.  
 
\section{\label{sec:04} Potential Tubes and Shells}

Everything was said about the volume capture in the section \ref{sec:02} can be apply to thin tubes or shells. Indeed, the volume potential in the Figs.\ref{fig:CurveSurface}, \ref{fig:Ellipse} does not effect the process of full internal reflection and volume capture. It can be cut off up to the sizes of the channels without any influence on channeling mode. The result will be the potential shells and closed tubes if 3D potentials are considered. If we consider the nonchanneled particles scattering on the bent shells and tube , the finite cross size of  shells and tubes can cause some interested effects noted in Fig. \ref{fig:ShiftRotation}.

It may be noted here that the effective cross-section of volume capture 
is independent of the sign of potential $U_1$ if geometric parameters (curvatures) of the potential surfaces are the same. If the potential is positive, we may consider external wall of this potential which acts in the same manner as a negative potential.  
So that the result is equally valid for repulsive and attractive fields.

\section{\label{sec:05} Conclusion}

Formulae (\ref{AngleRelationGeneral}), (\ref{crAngle0}) and (\ref{ImpactParamG}) give the general solution of the problem. The latter formula gives the relation between impact parameter of capture and curvature, i.e. the condition of volume capture in potential structures with variable curvature. Formula (\ref{crAngle0}) gives the relation between two impact angles inside the bent channel as an implicit function of integral curvature. 

To avoid any misunderstanding, of course, it should be noted that the way of expressing the real potential in the crystals through the uniform potential well is somewhat simplified picture. The real  crystals potential would be used for detailed calculations, and it may modify the quantitative results, but some  qualitative conclusions will definitely stay unchanged.

\bibliographystyle{elsart-num}
\bibliography{../../Focusing_and_Channeling_in_Crystals/chan02}%\bibliography{../../Focusing_and_Channeling_in_Crystals/math01}% Produces the bibliography via BibTeX.

\end{document}